\documentclass[letter,twocolumn]{jpsj2} 
\usepackage{dcolumn}
\usepackage{bm}

\title{Photoinduced metallic properties of one-dimensional strongly correlated electron systems}
\author{
Nobuya \textsc{Maeshima}${}^{1,2}$\thanks{E-mail: maeshima@ims.ac.jp} and
Kenji \textsc{Yonemitsu}${}^{2,3}$
}
\inst{
${}^1$ Department of Chemistry, Tohoku University, Aramaki, Aoba-ku, Sendai 980-8578\\
${}^2$ Institute for Molecular Science, Okazaki 444-8585 \\
${}^3$ Graduate University for Advanced Studies, Okazaki 444-8585 
}

\recdate{\today}

\abst{
We study photoinduced optical responses of one-dimensional strongly correlated electron systems. The optical conductivity spectra are calculated for the ground state and a photoexcited state in the one-dimensional Hubbard model at half filling by using the exact diagonalization method. 
It is found that, in the Mott insulator phase,  the photoexcited state has large spectral weights including the Drude weight below the optical gap.
As a consequence, the spectral weight above the optical gap is largely reduced.
These results imply that a metallic state is induced by photoexcitation.
Comparison between the photoexcited and hole-doped states shows that the photoexcitation is similar to chemical doping.
}

\kword{photoinduced phase transition, Mott insulator, Hubbard model}
\begin{document}
\sloppy
\maketitle


Since the discovery of high-T$_{\rm c}$ superconductivity in the carrier-doped copper oxides,~\cite{bednorz} doping effects in strongly correlated electron systems have been one of the most important subjects in the condensed matter physics.~\cite{imada}
In many cases, carrier doping is carried out chemically, for example, by the substitution of Sr for La in La$_2$CuO$_4$.~\cite{bednorz}  In addition to such chemical doping, photoirradiation is another way to introduce carriers into materials (photodoping).~\cite{yu} Photodoping effects have been studied in many strongly correlated electron systems.~\cite{koshi1,koshi2,tajima,iwai}

One-dimensional (1D) halogen-bridged Ni complexes (Ni-X chains) are also known as strongly correlated electron systems.~\cite{okamoto}  They are in the Mott insulator phase because of large on-site Coulomb interaction on Ni sites. Recently, in Ni-X chains, photoinduced enhancement of a Drude-like low-energy component has been found just after the photoirradiation, suggesting the Mott transition by photodoping.~\cite{iwai}


It is well known that chemical doping of Mott insulators enhances the Drude weight in optical conductivity,~\cite{uchida} and many theoretical studies on this effect have been performed, for example,  on the basis of the numerical diagonalization.~\cite{dagotto,dagotto2,eskes,nakano,tohyama}  By contrast, the photodoping effect on optical responses in Mott insulators has not been studied so intensively.~\cite{gomi}  Thus there remain several points unsolved.
First, it is unclear whether photoexcited states are really metallic states accompanied with a completely different electronic structure from that of Mott insulators or not. 
 It is established that Mott insulators have a characteristic electronic structure, the so-called upper Hubbard band and the lower Hubbard band.~\cite{hubbard}  Is such an electronic structure destroyed by photoexcitation?  A second point is on the relation between photodoped systems and chemically doped systems. In both systems, a number of carriers have significant influences on the physical properties. Thus we want to clarify how the photodoped systems are similar to chemically doped systems.

To address these issues, we study the optical responses of the photoexcited state in the 1D Hubbard model defined by
\begin{equation}
{\cal H} = -t\sum_{l,\sigma}( c^\dagger_{l+1,\sigma}c_{l,\sigma} +  c^\dagger_{l,\sigma}c_{l+1,\sigma} )
  + U\sum_l n_{l,\uparrow}n_{l,\downarrow} ,
 \label{eq:ham}
\end{equation}
where $c^{\dagger}_{l,\sigma}$ ($c_{l,\sigma}$) is the creation (annihilation) operator of an electron with spin $\sigma$ at site $l$, $n_{l,\sigma}=c^{\dagger}_{l,\sigma}c_{l,\sigma}$, and $n_{l}=n_{l,\uparrow}+n_{l,\downarrow}$.  The parameter $t$ denotes the nearest-neighbor transfer integral and $U$ the on-site repulsion strength. In what follows, $t$ is set to unity for simplicity.
 By using the exact diagonalization method, we calculate the optical conductivity and the Drude weight of the ground and photoexcited states.  Our results show that the photoexcited state is metallic with a large Drude weight and a significant spectral weight below the optical gap $\Delta_{\rm opt}$ shifted from the holon-doublon continuum.  As a consequence, the spectral weight above $\Delta_{\rm opt}$ is largely reduced. It is also found that the optical conductivity of the photodoped system is similar to that of a hole-doped system.


The optical conductivity of the ground state  $|\psi_{0}\rangle$ is given by
\begin{equation}
\sigma(\omega)\equiv D\delta(\omega) +\sigma^{\rm reg}(\omega),
\end{equation}
where $D$ is its Drude weight defined by
\begin{equation}
D = - \frac{\pi}{N} \langle\psi_0|\hat{K}|\psi_0\rangle - \frac{2\pi}{N}\sum_{n > 0}\frac{|\langle \psi_{ n}|\hat{j}|\psi_0\rangle|^2}{E_{n}-E_0},
\end{equation}
with $\hat{K}$ being the kinetic term of the Hamiltonian~(\ref{eq:ham}), $\hat{j}$ the current operator defined by $\hat{j}\equiv i\sum_{l,\sigma}( c^\dagger_{l+1,\sigma}c_{l,\sigma} -  c^\dagger_{l,\sigma}c_{l+1,\sigma})$, $E_0$ the ground state energy, $|\psi_n\rangle$ the $n$-th excited state, and $E_n$ the corresponding energy.
The regular component $\sigma^{\rm reg}(\omega)$ is  defined by
\begin{equation}
\sigma^{\rm reg}(\omega)= -\frac{1}{N\omega}{\rm Im}\left[\langle\psi_{\rm 0}|\hat{j}\frac{1}{\omega+i\epsilon+E_{\rm 0} -{\cal H}}\hat{j}|\psi_{\rm 0}\rangle \right],
\label{eq:sigma-reg}
\end{equation}
where $\epsilon$ gives a finite broadening and is set at 0.1 in our calculations below.

The optical conductivity of a photoexcited state is a key quantity.  Here, we need to specify a photoexcited state. Experimentally, photoexcited states are created by the irradiation of a femtosecond-pulse laser, and they relax to the ground state within a few picoseconds.~\cite{iwai,takahashi}  Hence the true photoexcited states are non-equilibrium states, whose optical conductivity is generally difficult to be evaluated even numerically.  Thus we here introduce some approximation on the photoexcited state: we regard the photoexcited state as the optically allowed, first excited state $|\psi_{\rm 1opt}\rangle$. In other words, $|\psi_{\rm 1opt}\rangle$ corresponds to the lowest peak of $\sigma^{\rm reg}(\omega)$.  This assumption would be valid when we treat the system just after the irradiation of a pulse laser of energy equal to $\Delta_{\rm opt}$.  Then the optical conductivity of $|\psi_{\rm 1opt}\rangle$ with energy $E_{\rm 1opt}$ is given by
\begin{equation}
\sigma_1(\omega)\equiv D_1\delta(\omega)  +\sigma_1^{\rm reg}(\omega)  +\sigma_1^{\rm reg}{'}(\omega),
\label{eq:sigma1}
\end{equation}
with
\begin{eqnarray}
&&\sigma_1^{\rm reg}(\omega) \equiv    \nonumber \\
 &-&\frac{1}{N\omega}{\rm Im}\left[\langle\psi_{\rm 1opt}|\hat{j}\frac{1}{\omega+i\epsilon+E_{\rm 1opt} -{\cal H}}\hat{j}|\psi_{\rm 1opt}\rangle \right], \ \ \ \ \ \ 
\end{eqnarray}
and
\begin{eqnarray}
&&\sigma_1^{\rm reg}{'}(\omega)  \equiv \ \ \ \ \ \ \ \ \ \nonumber \\
& &\frac{1}{N\omega}{\rm Im}\left[\langle\psi_{\rm 1opt}|\hat{j}\frac{1}{\omega+i\epsilon-E_{\rm 1opt} +{\cal H}}\hat{j}|\psi_{\rm 1opt}\rangle\right].\ \ \ \ \ 
\end{eqnarray}
Here $D_1$ is the Drude weight of $|\psi_{\rm 1opt}\rangle$ defined by
\begin{equation}
D_1 = - \frac{\pi}{N} \langle\psi_{\rm 1opt}|\hat{K}|\psi_{\rm 1opt}\rangle - \frac{2\pi}{N}\sum_{n\ne \rm 1opt}\frac{|\langle \psi_{ n}|\hat{j}|\psi_{\rm 1opt}\rangle|^2}{E_{n}-E_{\rm 1opt}}.
\end{equation}
We treat the half-filled $N$-site chains and impose the periodic boundary condition for $N$=$4n$ and the anti-periodic boundary condition for $N$=$4n$+$2$. In what follows, the system size is set at $N=12$ unless otherwise noted.  The Lanczos method is used for diagonalization.



\begin{figure}[hbt]
\begin{center}
\includegraphics[width=6.0cm,clip]{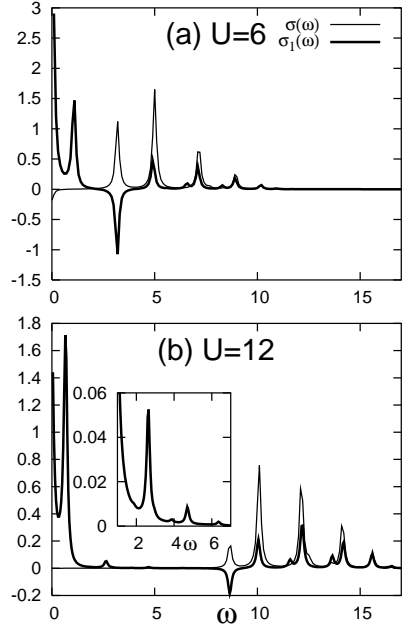}
\end{center}
\caption{Optical conductivity in the 1D Hubbard model, (a) for $U=6$, and (b) for $U=12$.
The thin (thick) lines show $\sigma(\omega)$ $\left[ \sigma_1(\omega) \right]$. The mid-gap region is magnified in the inset.}
\label{fig:misw}
\end{figure}

Figure~\ref{fig:misw} shows the optical conductivity $\sigma(\omega)$ and $\sigma_1(\omega)$ in the model~(\ref{eq:ham}).
In $\sigma(\omega)$, the spectral weight distributed over a frequency range from approximately $U-4t$ to $U+4t$ is due to the single holon-doublon pair excitations~\cite{stephan} and called the holon-doublon continuum hereafter. 
   Below $\Delta_{\rm opt}$, no spectral weight is observed except for the small negative $D$, which is visible only for $U=6$. The appearance of a finite $D$ is due to the finite-size effect, and $D$ exponentially decreases with $N$.~\cite{fye}

In $\sigma_1(\omega)$, by contrast, several peaks appear below $\Delta_{\rm opt}$. The most pronounced one is the Drude peak at $\omega=0$, implying the conducting property of the photoexcited state $|\psi_{\rm 1opt}\rangle$.  Another pronounced peak (we call it ``A'' below) is observed around $\omega \sim 1$.  The peak ``A'' is caused by the transition from $|\psi_{\rm 1opt}\rangle$ to the nearly degenerate state inherent in the 1D Mott insulators.~\cite{mizuno} 
Other small peaks are found below $\Delta_{\rm opt}$ [see the inset of Fig.~\ref{fig:misw}(b) ]. These are attributed to the transitions from $|\psi_{\rm 1opt}\rangle$, which is regarded as a holon-doublon pair excited state, to other holon-doublon pair excited states.   The negative peak at $\omega = \Delta_{\rm opt}$ corresponds to the transition from $|\psi_{\rm 1opt}\rangle$ to $|\psi_0\rangle$.  For $\omega > \Delta_{\rm opt}$, the continuum is somewhat similar to that in $\sigma(\omega)$, but its weight is reduced.  In particular, the decrease of the peak height becomes significant as the peak energy is lowered, which is consistent with the experimental observation in the Ni-X chains.~\cite{iwai}

\begin{figure}[hbt]
\begin{center}
\includegraphics[width=6.0cm,clip]{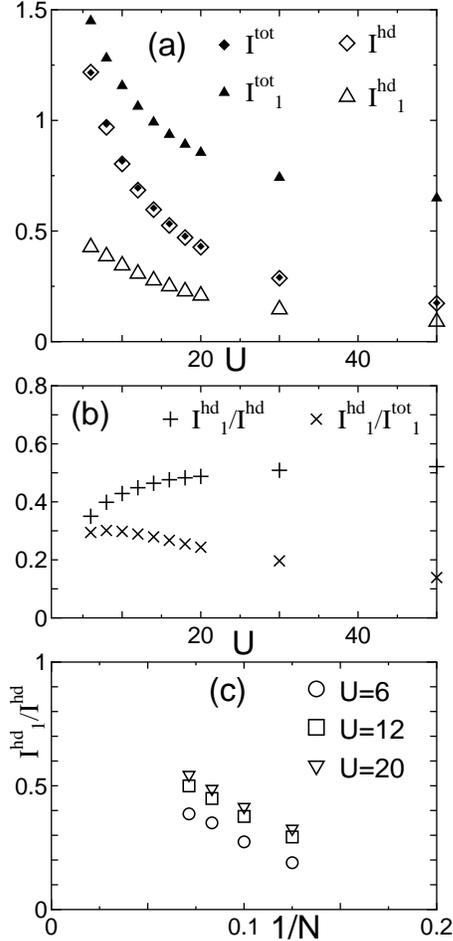}
\end{center}
\caption{ 
(a) Integrated intensities~(\ref{eq:Ihd})-(\ref{eq:I1tot}), and (b) their ratios in the 1D Hubbard model, as a function of $U$.
(c) shows the size dependence.
}
\label{fig:intsw}
\end{figure}

To clarify how the weight of the continuum ranging from $U-4t$ to $U+4t$ is reduced by photoexcitation, we calculate the integrated intensity defined by
\begin{equation}
{\rm I^{hd}} \equiv \int_{\omega_{\rm l}}^{\omega_{\rm u}}\sigma^{\rm reg}(\omega) d\omega 
\label{eq:Ihd}
\end{equation}
and
\begin{equation}
{\rm I^{hd}_1} \equiv \int_{\omega_{\rm l}}^{\omega_{\rm u}}\sigma_1^{\rm reg}(\omega) d\omega,
\label{eq:I1hd}
\end{equation}
where we set $\omega_{\rm l}$ and $\omega_{\rm u}$ such that these integrations only include the contribution from the holon-doublon continuum above $\Delta_{\rm opt}$.  We also calculate the total integrated intensity
\begin{equation}
{\rm I^{tot}} \equiv\int_0^{\infty}\sigma(\omega)d\omega  = \frac{\pi}{2N}\langle\psi_0|\hat{K}|\psi_0\rangle
\label{eq:Itot}
\end{equation}
and that of $\sigma_1(\omega)$,
\begin{equation}
{\rm I^{tot}_1} \equiv\int_0^{\infty}\sigma_1(\omega)d\omega  = \frac{\pi}{2N}\langle\psi_{\rm 1opt}|\hat{K}|\psi_{\rm 1opt}\rangle.
\label{eq:I1tot}
\end{equation}
These quantities are plotted in Fig.~\ref{fig:intsw}.
We confirm that, in $\sigma_1(\omega)$, the integrated intensity from the holon-doublon continuum above $\Delta_{\rm opt}$ is largely reduced: the ratio $I_1^{\rm hd}/I^{\rm hd}$ is about 1/3 for $U=6$. $I_1^{\rm hd}/I^{\rm hd}$ increases with $U$, suggesting that the holon-doublon continuum becomes robust as $U$ increases.  It should be noted that $I_1^{\rm hd}/I^{\rm hd}$ depends on the system size [Fig.~\ref{fig:intsw}(c)]. $I_1^{\rm hd}/I^{\rm hd}$ increases with $N$ owing to the decreasing density of photodoped carriers $\rho_{\rm PC}$. Because $|\psi_{\rm 1opt}\rangle$ has two carriers (one holon and one doublon), $\rho_{\rm PC}$ is considered to be $2/N$. Hence the influence of photodoping becomes weak as $N$ increases.  Note the large difference between $I_1^{\rm hd}$ and $I_1^{\rm tot}$.  In $\sigma_1(\omega)$, $I^{\rm hd}_1$ is much smaller than $I^{\rm tot}_1$.
Thus the low-energy components have dominant contribution to $\sigma_1(\omega)$. 
In $\sigma(\omega)$, almost all the intensity is contributed from the holon-doublon continuum.

\begin{figure}[hbt]
\begin{center}
\includegraphics[width=6.0cm,clip]{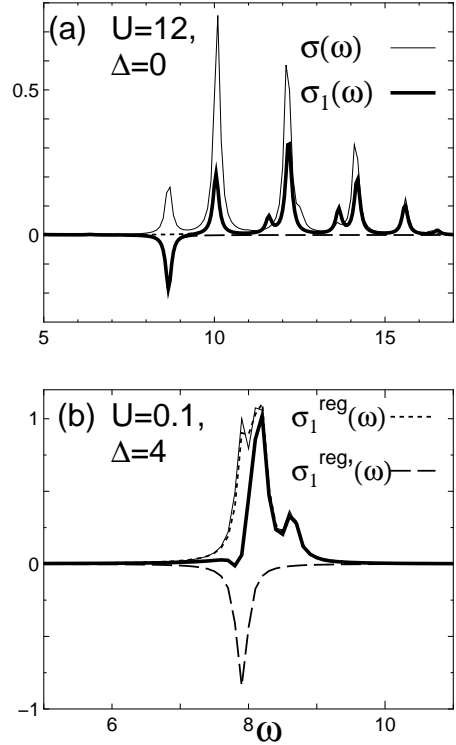}
\end{center}
\caption{Optical conductivity in the 1D IHM, (a) in the Mott insulator phase ($U=12$, $\Delta=0.0$), and (b) in the band insulator phase ($U=0.1$, $\Delta=4$).}
\label{fig:BIsw}
\end{figure}

The large shift of the spectral weight from above $\Delta_{\rm opt}$ to low energies is the characteristic doping effect in the Mott insulator phase.~\cite{uchida}  To confirm this in the photodoping case, we compare the results in the Mott insulator phase with those in the band insulator phase.  For this purpose we hereafter employ the 1D ionic Hubbard model (IHM) defined by
\begin{equation}
{\cal H}_{\rm IHM} = {\cal H} + \Delta \sum_l (-1)^l n_l,
\end{equation} 
where $\Delta(>0)$ denotes a half of the level difference between the neighboring orbitals.  This model can describe both the Mott insulator phase ($U\gtrsim\Delta$) and the band insulator phase ($U\lesssim\Delta$).~\cite{nagaosa,fabri}
   Figure~\ref{fig:BIsw} shows $\sigma(\omega)$ and $\sigma_1(\omega)$ in the Mott insulator phase ($U=12$,$\Delta=0$) and in the band insulator phase ($U=0.1$, $\Delta=4$).
In $\sigma_1(\omega)$ of the band insulator phase, the contribution from the hole-electron excitations~\cite{gebhard} is similar to that in $\sigma(\omega)$ except for the absence of the lowest peak around $\omega = 7.9(=\Delta_{\rm opt})$.
This single difference between $\sigma(\omega)$ and $\sigma_1(\omega)$  comes from the addition of negative $\sigma_1^{\rm reg}{'}(\omega)$ to $\sigma_1^{\rm reg}(\omega)$ in eq.~(\ref{eq:sigma1}).
  The bare contribution from the hole-electron pair excitation $\sigma_1^{\rm reg}(\omega)$ is almost the same as $\sigma(\omega)$, i.e., the peak positions and their heights are almost the same, suggesting that the band structure is almost unchanged by the photoexcitation.  This is in contrast to the situation in the Mott insulator phase.

We have discussed the optical responses of the photoexcited state, and found that, in the Mott insulator phase, the photodoping substantially shifts a spectral weight from above $\Delta_{\rm opt}$ to low energies including the Drude component.  We now turn to a chemically doped system for comparison with the photodoped system and discuss their similarities and differences. 
The system doped with two holes is treated to be compared with the photodoped system with one holon and one doublon.
 We calculate the optical conductivity
\begin{equation}
\sigma_{\rm 2h}(\omega)\equiv D_{\rm 2h}\delta(\omega) +\sigma_{\rm 2h}^{\rm reg}(\omega),
\end{equation}
where $D_{\rm 2h}$ is the Drude weight 
and $\sigma_{\rm 2h}^{\rm reg}$ is the regular component 
of the ground state $|\psi_{\rm 2h}\rangle$ of the two-hole-doped system.
The integrated intensities are also calculated, which are defined by
\begin{equation}
{\rm I^{hd}_{2h}}  \equiv \int_{\omega_{\rm l}}^{\omega_{\rm u}}\sigma_{\rm 2h}^{\rm reg}(\omega)d\omega,
\label{eq:I2hhd} 
\end{equation}
and
\begin{equation}
{\rm I^{low}_{1(2h)}} \equiv \int_0^{\omega_{\rm l}} \sigma_{\rm 1(2h)}(\omega)d\omega.
\label{eq:I2hlow}
\end{equation}

\begin{figure}[hbt]
\begin{center}
\includegraphics[width=6.0cm,clip]{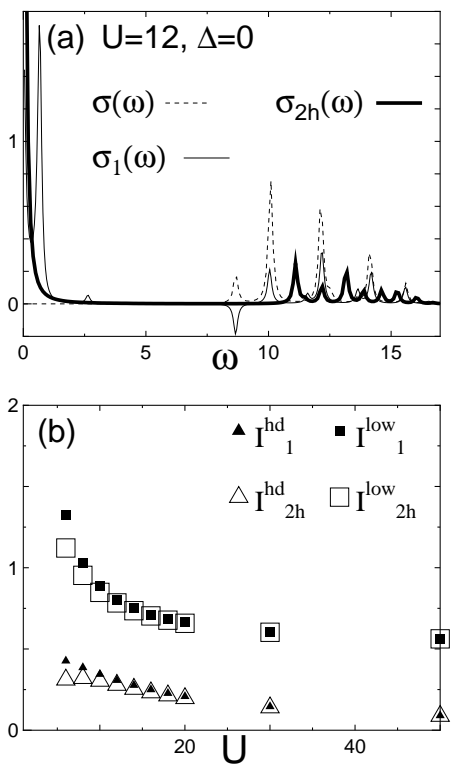}
\end{center}
\caption{
(a) Optical conductivity of the undoped (dashed line), photodoped (thin line), and 2-hole doped (thick line) 1D Hubbard model.
(b) Integrated intensities (\ref{eq:I1hd}), (\ref{eq:I2hhd}), and (\ref{eq:I2hlow}).
}
\label{fig:chemdope}
\end{figure}

It is found that both doped systems have the reduced holon-doublon continuum above $\Delta_{\rm opt}$ [Fig.~\ref{fig:chemdope}(a)].
The integrated weight $\rm I^{hd}_{2h}$ is comparable with ${\rm I^{hd}_{1}}$, and it approaches ${\rm I^{hd}_{1}}$ as $U$ increases [Fig.~\ref{fig:chemdope}(b)].
In the low-energy part, there exists some difference.  Although both systems exhibit a large Drude component, the hole-doped system has no analogue of the peak ``A'' in $\sigma_1(\omega)$.  In fact, $D_1$ is smaller than $D_{\rm 2h}$ roughly by the amount of the peak ``A'', so that the integrated low-energy weight $\rm I^{low}_{1}$ is comparable  with  $\rm I^{low}_{2h}$.

To summarize, we have studied the photodoping effect on the 1D Mott insulator phase by using the exact diagonalization method.
The optical conductivity shows that the photoexcited state in the Mott insulator phase has a large Drude weight, implying the metallic property. As a consequence, the contribution from the holon-doublon continuum above the optical gap, characterizing the 1D Mott insulators, is largely reduced, suggesting the large deformation of the electronic structure in the Mott insulator phase. 
The comparison with the chemically doped system shows the similarity between these doped systems.
This fact would be modified if the oppositely charged carriers are attracted to each other by introduction of the nearest-neighbor repulsion.

\section*{Acknowledgments}

The authors are grateful to H. Matsuzaki and H. Okamoto for showing their 
data prior to publication and for enlightening discussions.
This work was supported by Grants-in-Aid for Creative Scientific Research 
(No. 15GS0216), for Scientific Research on Priority Area ``Molecular 
Conductors'' (No. 15073224), for Scientific Research (C) (No. 15540354), and 
NAREGI Nanoscience Project from the Ministry of Education, Culture, Sports, 
Science and Technology, Japan.


\end{document}